\documentclass[12pt,preprint]{aastex}

\shorttitle{Dynamics and Metallicity of Bo\"{o}tes}
\shortauthors{Mu\~noz et al.}

\begin{document}

\title{Exploring Halo Substructure with Giant Stars: The Dynamics and Metallicity 
of the Dwarf Spheroidal in Bo\"{o}tes}

\author{
Ricardo R. Mu\~noz\altaffilmark{1},
Jeffrey L. Carlin\altaffilmark{1},
Peter M. Frinchaboy\altaffilmark{1},
David L. Nidever\altaffilmark{1},
Steven R. Majewski\altaffilmark{1} \&
Richard J. Patterson\altaffilmark{1}}

\altaffiltext{1}{Dept. of Astronomy, University of Virginia, P. O. Box
400325, Charlottesville, VA 22904-4325 (rrm8f, jc4qn, pmf8b, dln5q, srm4n,
ricky@virginia.edu)}

\begin{abstract}

We report the results of a spectroscopic study of the Bo\"{o}tes (Boo)
dwarf spheroidal (dSph) galaxy carried out with the WIYN telescope and 
the Hydra multifiber spectrograph. Radial velocities have been
measured for 58 Boo candidate stars selected to have magnitudes and
colors consistent with its red and asymptotic giant branches.  Within
the 13\arcmin~half-light radius, seven members of Boo yield a systemic
velocity of $V_{r}=95.6\pm3.4$ km s$^{-1}$ and a velocity
dispersion of $\sigma_{\rm o}=6.6\pm2.3$ km s$^{-1}$.  This implies a
mass on the order of $1\times10^7$ M$_{\sun}$, similar to the inferred
masses of other Galactic dSphs. Adopting a total Boo luminosity of
$L=1.8\times10^4$ L$_{\sun}$ to $8.6\times10^4$ L$_{\sun}$ implies
$M/L\sim610$ to 130, making Boo, the most distorted known Milky Way
dwarf galaxy, potentially also the darkest.  From the
spectra of Boo member stars we estimate its metallicity to be [Fe/H]
$\sim-2.5$, which would make it the most metal poor dSph known to date.

\end{abstract}


\section{Introduction}

Since the early work of \citet{Aaronson1983} on the Draco system, dwarf
spheroidal (dSph) galaxies in the Local Group have been suspected to
be heavily dark matter (DM) dominated.  Analyses of dSph internal dynamics
under the assumption of virial equilibrium suggest mass contents far
exceeding those inferred from their luminosities, with central
mass-to-light ratios ($M/L$) ranging from a few to about a hundred (in
solar units).  Such $M/L$ imply that dSphs have the largest
DM fraction of all galaxy types in the
universe.  Yet, despite the wide range of inferred $M/L$ for dSphs,
their central velocity dispersions ($\sigma_{\rm o}$) and half-light
radii ($r_{\rm h}$) seem to be remarkably similar, $\sim7-10$ km
s$^{-1}$ and $\sim200$ pc respectively.  Using these values, typical total
dSph masses of a few times $10^7$ M$_{\sun}$ are inferred (e.g.,
\citealt{Mateo1998})\footnote{Recent kinematical studies of dSphs have
attempted to derive their mass content by modeling the shape of the
velocity dispersion profile (e.g., \citealt{Kleyna2002};
\citealt{Walker2006a,Walker2006b}), but these calculations do not take
into account likely tidal effects in the outskirts of dSphs, as pointed
out by \citet{Lokas2005,M05,M06,Westfall2006,Sohn2006}.}.  This mass
conspiracy seemingly extends to even the very low end of the galaxy
luminosity scale.  An early kinematical survey of the dSph in Ursa
Major (UMa; \citealt{Kleyna2005}), the faintest and most diffuse of
the known Milky Way (MW) dwarf satellites (\citealt[][hereafter {W05}]{W05}),
found $\sigma_{\rm o}$ = 9.3$^{+11.7}_{-1.2}$ km s$^{-1}$, which,
coupled with UMa's $r_{\rm h}$ of 250 pc, also results in a mass of
$\sim10^7$ M$_{\sun}$.

Recently, a new Galactic dSph candidate has been found in the
constellation of Bo\"{o}tes (\citealt[][hereafter {B06b}]{B06b})
during a search for halo substructure using Sloan
Digital Sky Survey (SDSS; \citealt{Abazajian2005}) data.  Along with UMa,
Bo\"{o}tes (Boo) is one of the faintest MW satellites found,
having (at a 60 kpc distance) an absolute magnitude of $M_{V}=-5.8$
(but maybe brighter; see \S 3.2).  In addition, the Boo dSph exhibits
the most irregular density contours of any Galactic dSph
(\citeauthor*{B06b}), which suggests that the satellite may be undergoing
tidal disruption.

In this Letter, we present the results of a spectroscopic survey of
Boo candidate stars. 
We have identified its kinematical signature, which allows us to
estimate its systemic velocity, mass and central density, as well as
its $M/L$.  We also use the spectral data to make a rough measurement
of the metallicity of Boo giant stars.

\section{Spectroscopic Observations and Data Reduction}

Though \citeauthor*{B06b} detected Boo as an overdensity in SDSS Data 
Release 5 (DR5), that dataset was
proprietary at the time we selected observing targets.  However, Boo
appears near the northern limit of the SDSS Data Release 4 (DR4,
\citealt{sdss_dr4}) as well.  Because the core of the Boo dSph is well
within the DR4 coverage area, we select Boo candidate stars in the
core and southward of declination $\sim$14.8\arcdeg~(Figure 1).

Spectroscopic follow-up of Boo red giant branch (RGB) and asymptotic giant branch 
targets with $g$ $<$ 19.0, as shown in Figure 2,
was carried out with the WIYN 3.5-m
Hydra multi-fiber spectrograph on UT 2006 May 6--7.  
Targets are distributed
throughout the $\sim$1-degree WIYN+Hydra field of view.
The 600@10.1
grating was used with the red fiber cable to yield a wavelength
coverage $\lambda$ = 4500--7200 \AA~with a spectral resolution of 2.80
\AA~per resolution element. 
To achieve adequate $S/N$ for velocity determination, we
observed one Hydra configuration containing 72 Boo candidate stars for
$4 \times 30$ min.  We also observed 9 radial velocity (RV) standards
covering spectral types F through M, each through multiple
fibers, yielding a total of 64 RV cross-correlation template spectra.

Preliminary processing used the {\it IRAF} CCDRED package, with
spectral reduction following the standard DOHYDRA routine.  RVs were
derived using the {\it IRAF} FXCOR task, with each
spectrum cross-correlated against all RV standard spectra. The regions
around the H$\alpha$, Mg triplet, and H$\beta$ lines were used for
cross-correlation. In the end, we measured reliable RVs for 58 of the
72 observed Boo candidates.
RV standards (observed multiple times) showed deviations 
from the published IAU values for $V_{r}$ of less than 1 km s$^{-1}$ on average. 

RV uncertainties were determined using the \citet{Vogt1995} method,
based on analysis of repeatedly observed standard-star
spectra.  This technique takes advantage of the fact that the
Tonry-Davis Ratio (TDR, \citealt{TonryDavis1979}) scales with $S/N$,
so that individual RV uncertainties may be found via: $Error(V_{r})
= \alpha/(1+TDR)$, where $\alpha$ is determined from the standards.
For our set of 64 standards, we measure $\alpha$ = 107, which yields
an average velocity uncertainty of $\sim$ 4.0 km s$^{-1}$ (TDR $>$ 25,
with typical $S/N$ of 15-20).

\section{Spectroscopic Results}

\subsection{Bootes Membership}

With our RV data alone, the signal of Boo stars is not clear.  To
improve the contrast of Boo stars with respect to MW contaminants, we
make use of the gravity sensitive MgH absorption feature near 5150
\AA~ to remove foreground Galactic dwarfs, as we have done
photometrically in our previous dSph studies (e.g.,
\citealt{Majewski2005}; \citealt{M05,M06}; \citealt{Westfall2006}).
Because the MgH absorption is very strong in all but the most
metal-poor ([Fe/H] $\lesssim$ -2) K--dwarf stars, visual inspection
alone is sufficient to ``clean" our spectroscopic sample of the
majority of foreground dwarfs by simply identifying those with strong
MgH features, like the one shown in Fig.\ 4 (see also Fig.\ 1 in
\citealt{Majewski2000} for an illustration).

To further clean this sample, we make use of instrumental Lick
spectroscopic indices (\citealt{Worthey1994}).  A proper determination
of stellar $T_{\rm eff}$, log $g$ and [Fe/H] values based on Lick
indices is beyond the purpose here.  Instead, we 
identify stars with similar metal line strengths using
the quantitative indices of the Lick system.
Figure 3a shows the Mg$_{1}$ + Mg$_{2}$ versus Mg
$b$ trend for all
stars visually classified as likely giants with 
RV uncertainties less than 7.5 km s$^{-1}$.  Figure 3b shows their RV
distribution as a function of radial distance from the nominal center
of Boo. We have marked those having relatively low and high Mg indices
with circles and squares respectively.  We also mark with asterisks a
group of stars clumped in $V_{r}$ ($\sim50$ km s$^{-1}$ in Fig.\
3b) and Mg strength.  These stars have
velocities compatible with those expected for debris from the leading
arm of the Sagittarius (Sgr) dSph (\citealt{Law2004}); because Boo
lies in the background of the Sgr leading arm (see Fig. 1 in
\citealt{field_of_streams2006}), giant stars with these RVs are not
unexpected in our sample.  Figure 3b shows that within the $r_{\rm h}$
of Boo the distribution of stars shows a distinct grouping of giant
stars with low Mg indices and $V_{r}\sim100$ km s$^{-1}$.
Such a velocity grouping is unexpected for a random halo population at
this Galactic position, and is clearly distinct from the distribution
of stars beyond $r_{\rm h}$, which is centered at 0 km s$^{-1}$ (the
nominal mean velocity for any MW stellar population at this Galactic
longitude).  A two-sample Kolmogorov-Smirnov test shows that the
populations inside and outside 13\arcmin~ are drawn from different
RV distributions with a probability greater than 99.99\%.  Thus, we
identify this group of RV-clumped stars within 13\arcmin~ with the Boo dSph.

This ``cleaner" RV distribution inside $r_{\rm h}$ helps us define a
membership criterion to identify Boo stars.  We first draw attention
to the broadening of the RV distribution beyond $\sim10$\arcmin.  This
behavior might be expected if Boo is undergoing tidal stripping, as
its distorted morphology suggests. Models of tidally disrupting
satellites predict a rising velocity dispersion at large radii (e.g.,
\citealt{Kroupa1997}), a trend that is also observed in a number of
other Galactic dSphs: Ursa Minor (UMi, \citealt{M05}), Sculptor
(\citealt{Westfall2006}), Carina (\citealt{M06}), Fornax
(\citealt{Walker2006a}), Sextans (\citealt{Walker2006b}) and Leo I
(\citealt{Sohn2006}).  Because of the broadening of the RV
distribution, we conservatively use the innermost seven stars to
define a 3$\sigma$ (see \S 3.2) RV criterion for membership of
$75<V_{r}<116$ km s$^{-1}$.  Twelve stars lie within this velocity
range, all with low Mg line strengths, the most widely separated
(within our survey) at 27\arcmin~ to the south of the Boo center.

Two likely giants at higher velocity than our 3$\sigma$ RV limit are
seen in Figure 3b.  They lie right in the narrow RGB of Boo (see
Fig. 2 where they are marked with open circles) and the strength of
their Mg features makes them consistent with being Boo members; these
stars suggest a possibly dramatic increase of the Boo velocity dispersion
with radius.

\subsection{Velocity Dispersion, Mass and $M/L$}

Using the innermost seven stars to define the properties of the Boo
core RV distribution gives a systemic velocity of
$V_{r}=95.6\pm3.4$ km s$^{-1}$ and a velocity dispersion of
$\sigma_{\rm o}=6.6\pm2.3$ km s$^{-1}$, calculated using the Maximum
Likelihood Method (\citealt{Pryor1993}; \citealt{Hargreaves1994};
\citealt{Kleyna2002}).  This $V_{r}$ translates into a
$V_{GSR}=102.9\pm2.8$ km s$^{-1}$, implying that Boo is likely not in
a circular orbit, but in a rather radial one.  Including all twelve
potential $3\sigma$ members in the calculation yields
$V_{r}=98.4\pm2.9$ km s$^{-1}$ and $\sigma_{\rm o}=9.0\pm2.2$ km
s$^{-1}$; the latter corroborating the apparent increase of the
velocity dispersion with radius.  If we include the two likely high
velocity members that lie above the 3$\sigma$ RV limit (\S 3.1), we
obtain $\sigma_{\rm o}=14.6\pm3.0$ km s$^{-1}$.

As is customary for this type of study, one can use the observed
$\sigma_{\rm o}$ to estimate the mass content in the Boo dSph under
the assumption that it is in dynamical equilibrium.  While this
assumption has been a matter of some debate, it has been shown (e.g.,
\citealt{PP95}; \citealt{OLA95}; \citealt{Kroupa1997}; Mu\~noz et al.,
in prep.) that unless the satellite is completely destroyed or near
complete destruction, the value of $\sigma_{\rm o}$ indeed reflects
the instantaneous mass content.  From \citet{Illingworth76}, the total
mass of the dSph system can be approximated by: $M_{\rm tot} = 167
\beta R_{c,g} V^{2}_{s}$, where $\beta$ is a correction parameter
dependent on the concentration value, $R_{c,g}$ is the geometric-mean
King core radius in pc, and $V_{\rm s}$ is a parameter related to
$\sigma_{\rm o}$.  \citet{Mateo1998} approximates $\beta=8.0$ and
$V_{s}\sim\sigma_{\rm o}$ for pressure-supported systems that follow
low concentration King models, as dSphs do.  We can further
approximate $R_{c,g}$ to $r_{\rm h}$ derived by \citeauthor*{B06b}
(13\arcmin, corresponding to 225 pc at a distance of 60 kpc).
With $\sigma_{\rm o}=6.6\pm2.3$ km s$^{-1}$ we obtain 
$M_{\rm tot} = 1.1^{+1.3}_{-0.5}\times10^7$ M$_{\sun}$.
We can also estimate the central mass density in Boo as 
$\rho_{\rm o} \approx 166 \sigma^{2}_{\rm o} \ R^{2}_{c,g}$ \citep{Mateo1998};
this yields $\rho_{\rm o} = 0.14 M_{\sun}$ pc$^{-3}$.

These results are remarkably similar to the values obtained for the
majority of the Galactic dSphs. \citet{Mateo1998} points out that
Local Group dSph systems seem to be embedded in DM halos of
$M_{\rm tot} \approx 10^7$ M$_{\sun}$ regardless of the contribution
of their luminous component. He further finds the following empirical
relation between the $M/L$ and the total luminosity of a dSph: $M/L =
2.5+[10^{7}/(L/L_{\sun})]$.  Adopting an absolute magnitude for Boo of
$M_{\rm V}=-5.8$ (from \citeauthor*{B06b}) yields a total luminosity
of $L_{\rm V} = 1.8\times10^{4}$ L$_{\sun}$, which, in turn, gives $M/L_{\rm
V}=610$ (M/L)$_{\sun}$. We note that this luminosity would make Boo
even fainter than UMa if we adopt for the latter the absolute magnitude
estimate of $M_{\rm V}=-6.75$ from \citeauthor*{W05}.
A visual comparison of the CMDs of Boo and UMa reveals that the Boo
RGB contains at least a factor of two more stars than that 
of UMa (see also \citealt{Siegel2006}).
Since they have comparable half-light radii, this implies that Boo
must be at least twice as bright as UMa. Were we to adopt the
\citeauthor*{W05} estimate for the UMa $M_{\rm V}$, this yields a
total luminosity for Boo of $L_{\rm V}\sim8.6\times10^{4}$ L$_{\sun}$,
making $M/L_{\rm V}=130$.  With either of these values, however, Boo
lies squarely on the \citet{Mateo1998} relation.

\subsection{Metallicity}

The Mg I triplet+MgH absorption features near 5150 \AA~ can be
exploited to estimate the Boo [Fe/H].  We have already shown (Fig.\ 3)
that Boo stars exhibit by far the weakest Mg features of all likely giant stars in
our sample.  We have
used the CTIO 4-m telescope + the Hydra multifiber spectrograph, with
a comparable instrument setup (yielding similar spectral resolution)
to that used for the Boo observations, for RV measurements of red
giant candidates in the globular clusters (GCs) NGC 288 and NGC 5634
(Moskowitz et al., in prep.).  These clusters have [Fe/H] of $-1.24$
and $-1.88$ respectively (\citealt{Harris1996}) and we use them as
metallicity calibrators, selecting for this purpose giant stars that
are confirmed RV cluster members.  In order to minimize the effect of
surface gravity and temperature in the measurements of equivalent
width (EW) of the Mg features, we pick and combine the spectra of stars
that lie in the upper part of their respective red giant branches,
which provides bright members with fairly similar colors. 
Figure 4 shows the combined spectra obtained for the two clusters and for Boo. 
We then add the EWs for the three Mg lines to compensate for the
weakness of the Mg features, and assume a linear 
function between this sum and [Fe/H]\footnote{Figure 9 from \citet{Buzzoni1992} 
suggests that a linear relationship between Mg$_{2}$ and [Fe/H] is not 
unreasonable for [Fe/H] $< -1.3$.}.
Using this derived relationship and assuming
a similar [Mg/Fe] ratio between the GC and Boo stars\footnote{\citet{Shetrone2001} and 
\citet{Shetrone2003} showed 
that metal poor stars in Sculptor, Leo I and UMi
([Fe/H] $<-2.0$) have comparable [$\alpha$/Fe] to Galactic GC stars.},
we translate the EWs measured for Boo into an extrapolated [Fe/H] of $-2.5$. 
The uncertainties in the EW measurements translate into
uncertainties in [Fe/H] of about 0.2 dex. This does
not include, of course, the error introduced by assuming a linear 
relationship between EWs and [Fe/H] in the first place, nor the uncertainties 
due to surface gravity, temperature effects, or possible variations in 
$\alpha$-element abundances between the GCs and Boo.

These calculations, although only intended to provide a very rough
estimation of the metallicity of Boo, are consistent 
with the [Fe/H] = $-2.6$  derived by \citet{Siegel2006} using RR Lyrae variables,
and with the fact that the Boo RGB seems to be slightly bluer
than that of M92 ([Fe/H] = $-2.3$; see Fig. 2 of \citeauthor*{B06b}).
Boo is potentially the most metal poor of the Galactic dSphs known to date. 

\section{Discussion}

We spectroscopically survey the Boo dSph and 
derive both its systemic velocity
($95.6\pm3.4$ km s$^{-1}$) and central velocity dispersion
($6.6\pm2.3$ km s$^{-1}$), which
yields a mass of $M_{\rm tot} = 1.1^{+1.3}_{-0.5}\times10^7$ M$_{\sun}$.
This mass is
similar to that of the other dSph galaxies and puts Boo squarely on the
``same mass-just different luminosities" trend identified by \citet{Mateo1998},
despite the fact that Boo is one of the faintest known Galactic satellites. 

The dynamical mass derived for Boo, taken at face value, implies that
this is possibly also the darkest dSph known to date.  If other
systems of similar luminosity have the same mass, the current pace of
discovery of these systems (three in the past year;
\citeauthor*{W05}; \citealt{Zucker_CVn2006}; \citeauthor*{B06b})
--- and in only the approximately 20\% of the sky covered by the SDSS
DR5 --- will help alleviate the current order of magnitude or two
deficit of known Galactic satellites compared to that predicted by
$\Lambda$CDM simulations, albeit only for one part of the mass
spectrum exhibiting the apparent ``missing satellites" shortfall
(e.g., \citealt{Klypin1999}).  While Boo seems
to be the most DM dominated dwarf, it is, at the same time,
the Galactic satellite with the most distorted known morphology
(\citeauthor*{B06b}) and possibly most dramatic increase in velocity
dispersion with radius.  This implies that Boo may be among the most
disrupted Galactic dwarfs, to the extent that
it even lacks a proper core (\citeauthor*{B06b}).  In fact a puzzling
correlation is now emerging between the DM fraction of a dSph
and its morphology, wherein the faintest and most distorted systems (the UMi, UMa and Boo dSphs)
seem also to present the largest central $M/L$'s.  If the distorted contours
are a response to the influence of Galactic tides, and if all dSphs indeed
have a similar current total mass and density, then why is it that tides 
seem to affect preferentially the least luminous systems?  
Perhaps these faintest systems
represent the \citet{Kroupa1997} regime in which disruption has
proceeded to the point where the central velocity dispersions
are inflated by tides, artificially increasing the derived $M/L$. But
even if so, then it is curious that the current
$\sigma_{\rm o}$ for Boo would be inflated to just such a value that the derived 
(but artificial) dynamical mass still participates in the ``same mass" 
conspiracy of dSph galaxies.

We appreciate useful discussions with Gregory Sivakoff, Jeffrey Crane and Allyson
Polak.  We gratefully acknowledge support by NSF grant AST-0307851,
NASA/JPL contract 1228235, the Virginia Space Grant Consortium and
Frank Levinson through the Celerity Foundation. 
D.L.N. is also supported by the ARCS Foundation and the Green Bank Telescope
Student Support Program. P.M.F. is supported by NASA GSRP and UVa dissertation fellowships.

\clearpage

\begin{figure}
\plotone{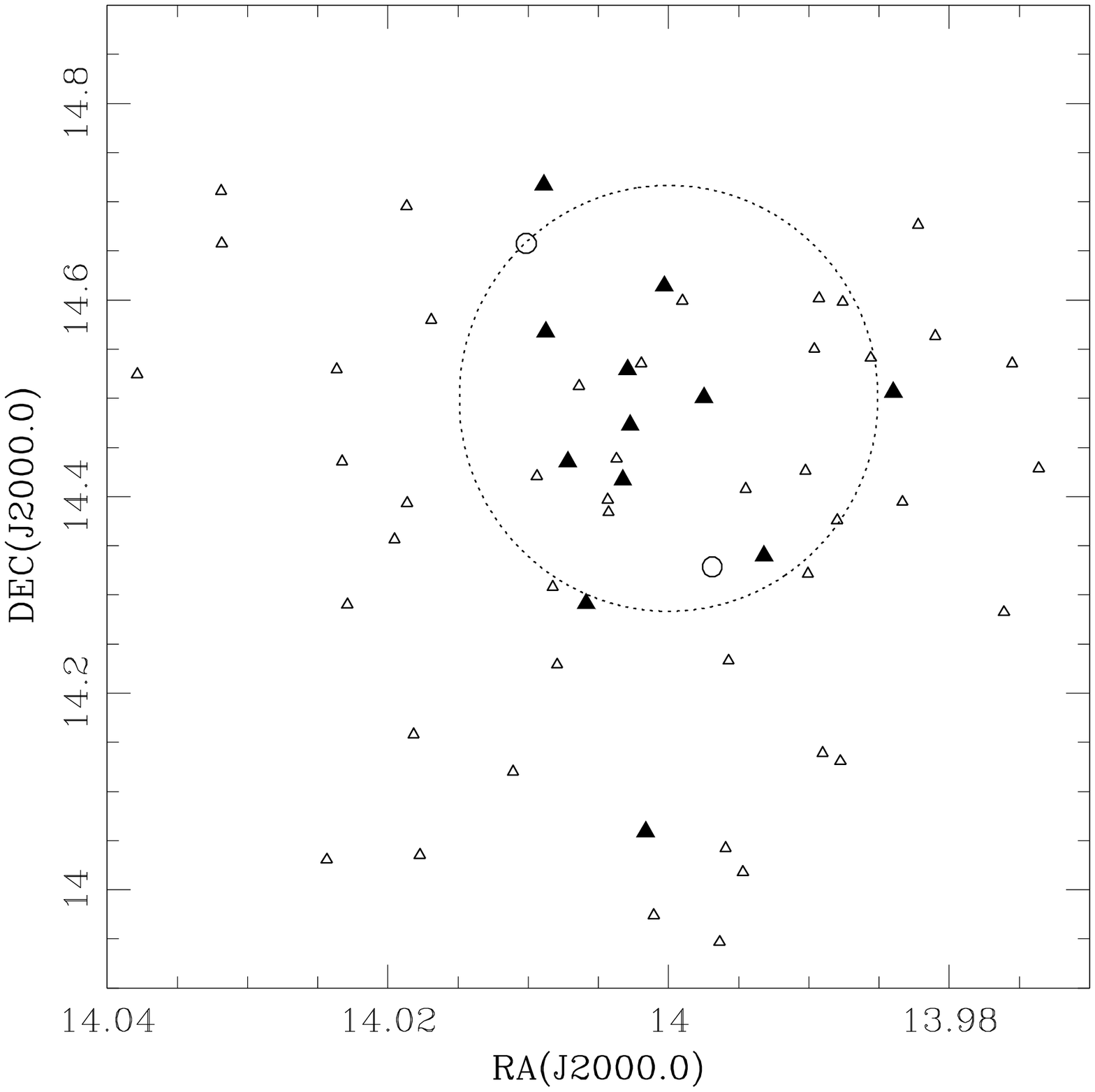}
\caption{Spatial distribution of all 58 stars with measured RVs.
Filled triangles correspond to the final Boo sample (from Fig.\ 3 analysis),
while open triangles mark stars not considered members.
Two open circles represent the higher-velocity stars discussed in the text as 
possible members based on the similarity of their Mg indices to those of other Boo stars.
The Boo $r_{\rm h}$ determined by \citeauthor*{B06b} is delineated by the dotted circle.}
\end{figure}

\clearpage

\begin{figure}
\plotone{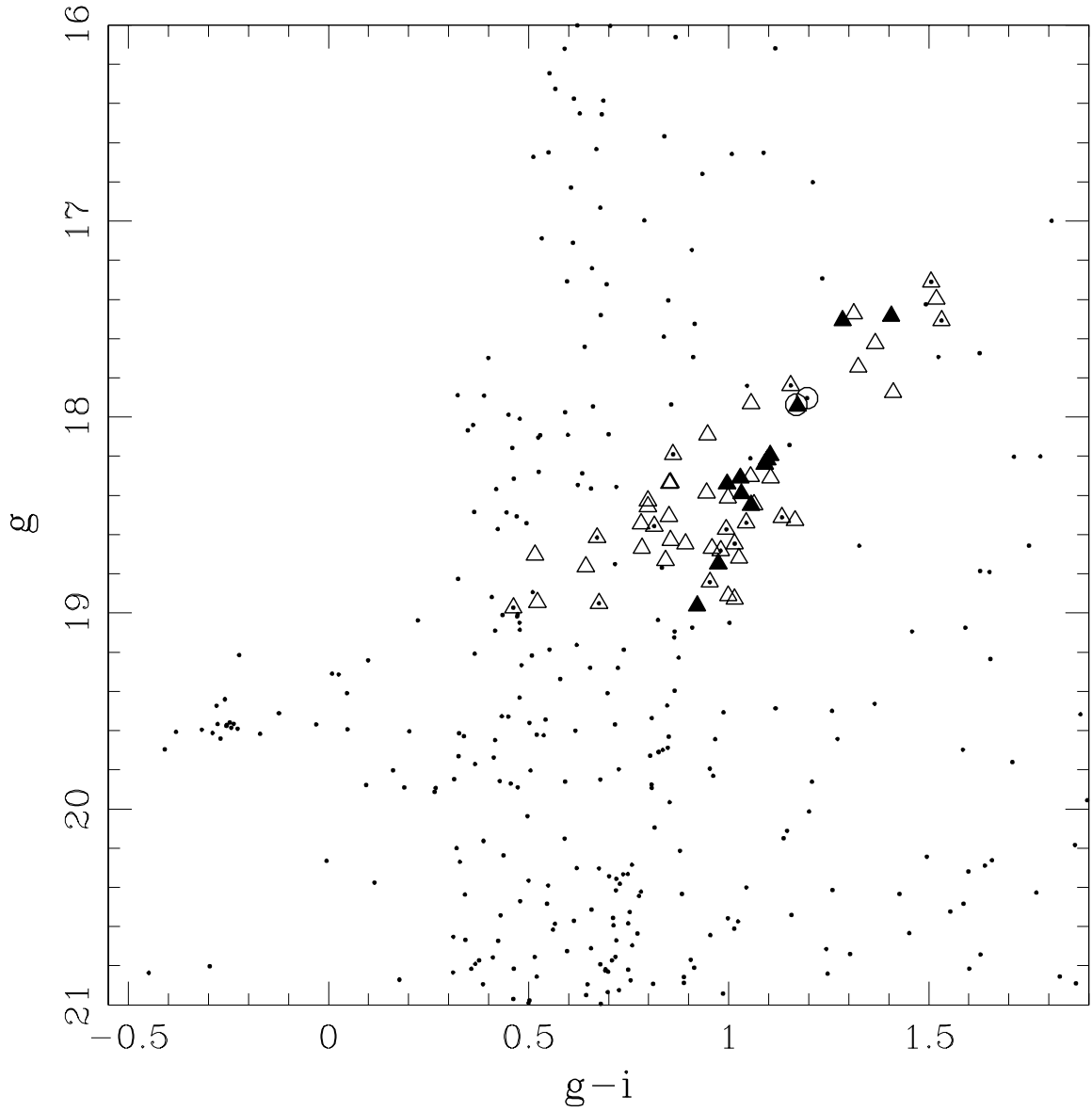}
\caption{CMD of the Boo region showing the clear Boo RGB and
horizontal branch.  The dots represent stars within the $r_{\rm h}$ of
Boo (dotted circle in Fig.\ 1) in order to highlight Boo's features,
while the other symbols show {\it all} 58 stars as in Fig.\ 1.}
\end{figure}

\clearpage

\begin{figure}
\plotone{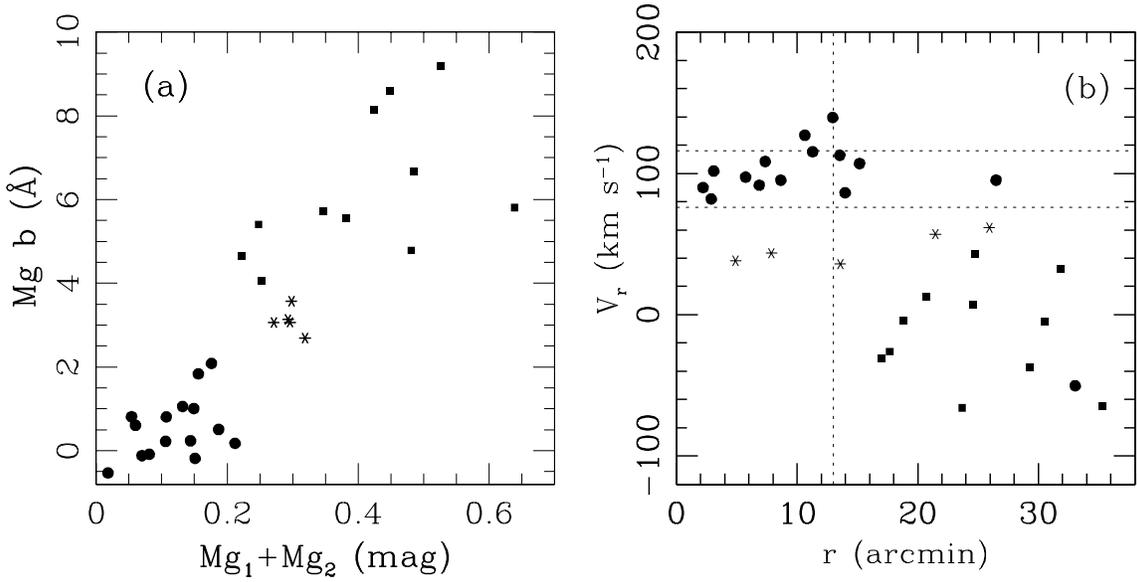}
\caption{(a) Mg$_{1}$+Mg$_{2}$ versus Mg $b$ instrumental Lick
indices for all stars with velocity uncertainties $<$ 7.5 km s$^{-1}$
visually classified as likely giants.  Circles mark stars most likely
to be metal poor, while squares show stars more likely to be metal
rich. We mark with asterisks a clump of stars with similar indices and
RVs, possibly from the Sgr dSph.  (b) RVs of all stars in (a) as a
function of radial distance from the center of the Boo dSph. Symbols
as in panel (a).  The dotted vertical line marks the Boo $r_{\rm h}$,
while the dotted horizontal lines delimit a 3$\sigma$ RV spread. The
two higher-velocity stars discussed in \S 3.1--3.2 (open circles in
Fig.\ 1) lie just above the upper 3$\sigma$ limit.}
\end{figure} 

\clearpage

\begin{figure}
\plotone{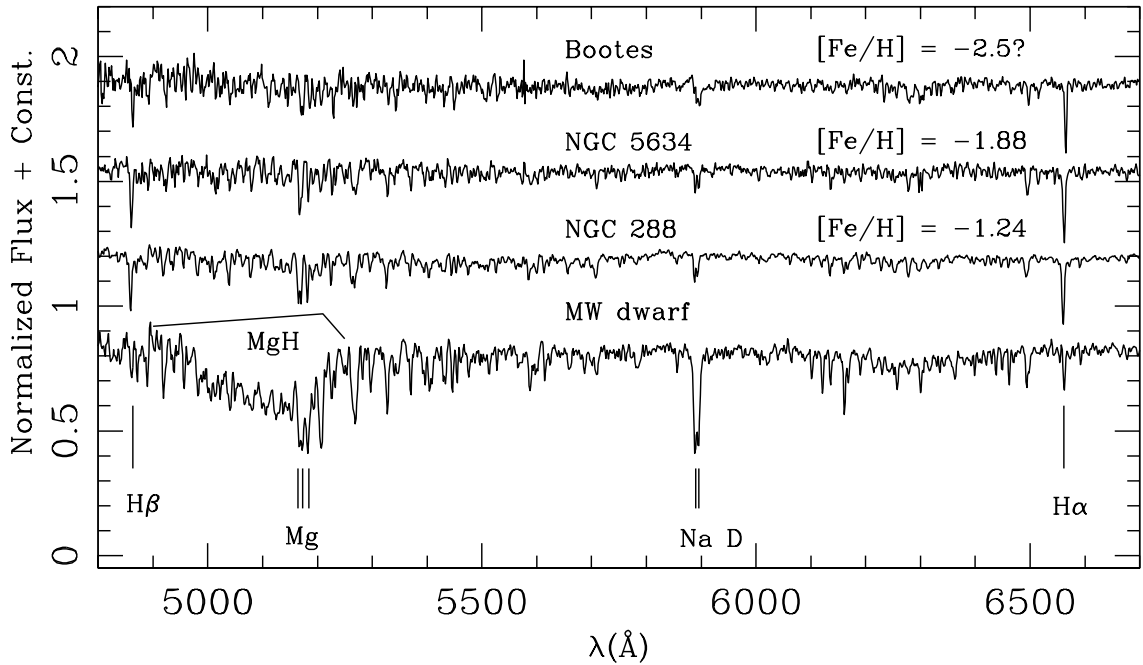}
\caption{Sample of the spectra used to estimate the metallicity of
Boo. From top to bottom: Combined spectrum of the three brightest Boo
giant stars, combined spectrum for NGC 5634 giants and combined
spectrum for NGC 288 giant stars.  The comparison spectrum of a dwarf star
demonstrates the broad MgH absorption feature between $\sim$4900-5250 \AA, which is an
obvious signature of late G or K dwarf stars.}
\end{figure}

\end{document}